\documentclass[12pt,twoside]{article}
\usepackage{fleqn,espcrc1}
\usepackage{graphicx,epsf}

\newcommand{\AmS}{{\protect\the\textfont2
  A\kern-.1667em\lower.5ex\hbox{M}\kern-.125emS}}

\title{Selected topics on  Low Energy Antiproton Physics}

\author{J. Carbonell\address
      {Institut des Sciences Nucl\'eaires, \\ 
        53, Av. des Martyrs, 38026 Grenoble, France}, M. Mangin-Brinet\addressmark}
       
\begin{document}

\maketitle

\begin{abstract}
Some of the last results on low energy antiproton physics are reviewed. 
First Faddeev calculations for \={n}d scattering length are presented.
\end{abstract}

%%%%%%%%%%%%%%%%%%%%%%%%%%%%%%%%%%%%%%%%%%%%%%%%%%%%%%%%%%%%%%%%%%%%%%%%%%%%%%%%%%%%%%%%%%%%%%
\section{Introduction}

We review in this contribution
some topics on the Low Energy Antiproton Physics
that raised our interest since the early shutdown of LEAR, 
as well as their relation with the theoretical models.

The first section is devoted to \={N}N system. 
We  comment on the \={n}p results
obtained by the OBELIX group
and discuss with some more details the protonium P-level
energy shifts measured by the PS207 experiment.

The second section is devoted to the \={N}d system.
We present the first Faddeev calculations for the \={n}d scattering length
and review the future perspectives from the theoretical side.
Unable to understand the annihilation at a quark level,
we have here the possibility to understand it at the level of nucleons.

%%%%%%%%%%%%%%%%%%%%%%%%%%%%%%%%%%%%%%%%%%%%%%%%%%%%%%%%%%%%%%%%%%%%%%%%%%%%%%%%%%%%%%%%%%%%%%%%%%%%
\section{Selected topics on \={N}N}

%%%%%%%%%%%%%%%%%%%%%%%%%%%%%%%%%%%%%%%%%%%%
\subsection{Low energy \={n}p cross section}

The last  OBELIX data on \={n}p cross section \cite{Iazzi_PLB_00}, 
presented in this conference by A. Felicello \cite{Felicello_LEAP_00}, are very astonishing. 
The structures observed in the total and elastic cross section 
at antineutron laboratory momenta $p_{lab}\approx70$ MeV/c  are not trivial results
and can hardly be explained without assuming a nearthreshold 
state \cite{Felicello_LEAP_98}.

These data have suffered
from big fluctuations in the successive steps of 
their analysis \cite{Iazzi_LEAP_92,Giacobbe_LEAP_96} and 
the only interesting question --  i.e. to what extend they are significant 
 -- can not longer be answered at LEAR.
The anomalous behaviour indeed concerns only two data points but if they were
a consequence of an experimental problem,
why should it manifest only in these two intermediate points ?
As strange as they could appear, these data
still remain inside the unitarity constraints.
It would be of high interest if the AD could allow low
energy scattering experiments.
If a direct \={n}p measurement turns out to be not possible,
one could see at least whether or not the structure
is manifested in the \={p}p cross section.

The problem is modelized by
\begin{equation}\label{CCOM}
(E-H_0)\pmatrix{\Psi_{\bar{p}p} \cr \Psi_{\bar{n}p} \cr \Psi_{\bar{n}n}} = \hat{V} \Psi
\qquad \hat{V}={1\over2}\pmatrix{V_0+V_1+2V_c&0&V_0-V_1\cr 0&V_1&0\cr V_0-V_1&0&V_0+V_1}
\end{equation} 
where $E=T_i+\sum m_i$ denotes the total energy -- the same for all channels,
$V_i$ the isospin components of the \={N}N strong potential and $V_c$
the Coulomb attraction.\
The \={n}p channel is dynamically decoupled from the two others.
However, through the T=1 component of the \={N}N potential, the \={n}p  structure 
could be visible in the \={p}p elastic cross section near the \={n}n threshold. 
Indeed, by assuming 
$2\mu_{\bar{p}p}=2\mu_{\bar{n}p}=2\mu_{\bar{n}n}={1\over2}(m_p+m_n)$
the center of mass momenta $k_i$ of the different channels are given by
\[ k_{\bar{p}p}^2 = k^2_{\bar{n}p}+\Delta= k^2_{\bar{n}n}+2\Delta\]
where $\Delta=m_n-m_p=1.293$ MeV.
The \={n}n threshold ($k^2_{\bar{n}n}$=0) correspond to the laboratory moment
$p_{\bar{p}p}$=98.6 and $p_{\bar{n}p}$=69.7 MeV/c, i.e. the region of the observed anomaly.
We would like to notice however that this coincidence is only kinematical.
Calculations performed in the framework of both optical and unitary coupled-channel models 
showed that the influence of opening the \={n}n threshold 
in the \={p}p cross sections indeed exists but it is negligible. If some
structure is seen in \={p}p observables near the \={n}n
threshold  it should have a dynamical origin.

%%%%%%%%%%%%%%%%%%%%%%%%%%%%%%%%%%%%%
\subsection{Protonium P-level shifts}

One of the most interesting results from LEAR {\it post-mortem} experiments  is 
the $^3$P$_0$ protonium level shifts obtained in  
\cite{G_NPA_99}. These authors found the values
$\Delta E_R({^3P_0})= -139\pm28$ and  $\Gamma({^3P_0})=120\pm 25$  meV.
The interest of this result is twofold. First 
because it is in qualitative agreement with a non trivial prediction
of some meson exchange based models. Second 
for the underlying  dynamics that it supports. 

\begin{table}[htb]
\caption{Protonium P-level shifts ($\Delta E_R-i{\Gamma\over2}$ in meV)
and \={p}p scattering volumes (a=$a_R+i\,a_I$ in fm$^3$) in some optical models.
The $^3P_0$ value is compared to experimental result from \cite{G_NPA_99} }
\label{POM}
\newcommand{\m}{\hphantom{$-$}}
\newcommand{\cc}[1]{\multicolumn{1}{c}{#1}}
\renewcommand{\tabcolsep}{0.65pc} % enlarge column spacing
\renewcommand{\arraystretch}{1.} % enlarge line spacing
\begin{tabular}{@{}c cc cc cc cc }\hline
          &  {\rm KW}    &                  & {\rm DR1}    &                  &  {\rm DR2}   &                  &    Exp       &                \\ \hline 
%\multicolumn{9}{c}{\hbox{Protonium level shifts (meV) }}  \\\hline  
State     & $\Delta E_R$ & ${\Gamma\over2}$ & $\Delta E_R$ & ${\Gamma\over2}$ & $\Delta E_R$ & ${\Gamma\over2}$ & $\Delta E_R$ &${\Gamma\over2}$\\ \hline 
$^1P_1$   &  -29.        & 13.              & -26.         & 13.		            & -24.         & 14.  	           &              &                \\ 
$^3P_0$   &  -69         & 48               &-74           & 57               & -62          &  40              & -139$\pm$28   &   60$\pm$13  \\
$ ^3P_1$  &  +29.        & 11.              & +36.         & 10.		            & +36.         &  8.8 	           &              &                \\
$^3PF_2$  &   -8.5       & 18.              & -4.8         & 15.		            & -5.9         & 16.  	           &              &                \\ \hline
%\multicolumn{9}{c}{\hbox{P-wave \={p}p scattering volumes a=$a_R$-$a_I\,$i (fm$^3$)  }}  \\ \hline  
          & $a_R$        & $a_I$            &  $a_R$       & $a_I$            &  $a_R$       &    $a_I$         &   $a_R$      &    $a_I$       \\ \hline
$^1P_1$   & -1.19        & -0.53            &  -1.07       & -0.52            & -0.99        & -0.58            &   	          &                \\
$^3P_0$   & -2.81        & -1.99            &  -3.01       & -2.31            & -2.53        & -1.62            &-5.68$\pm$1.14&-2.45$\pm$0.53 i \\  
$^3P_1$   & +1.22        & -0.47            &  +1.46       & -0.42            & +1.48        & -0.36	           &              &          \\
$^3PF_2$  & -0.36        & -0.75            & -0.20        & -0.63            & -0.25        & -0.67            &	             & \\\hline
\end{tabular} 
\end{table}

\begin{figure}[htb]
\vspace{-0.4cm}
\begin{minipage}[t]{75mm}
\hspace{0.5cm}\epsfxsize=7.cm\mbox{\epsffile{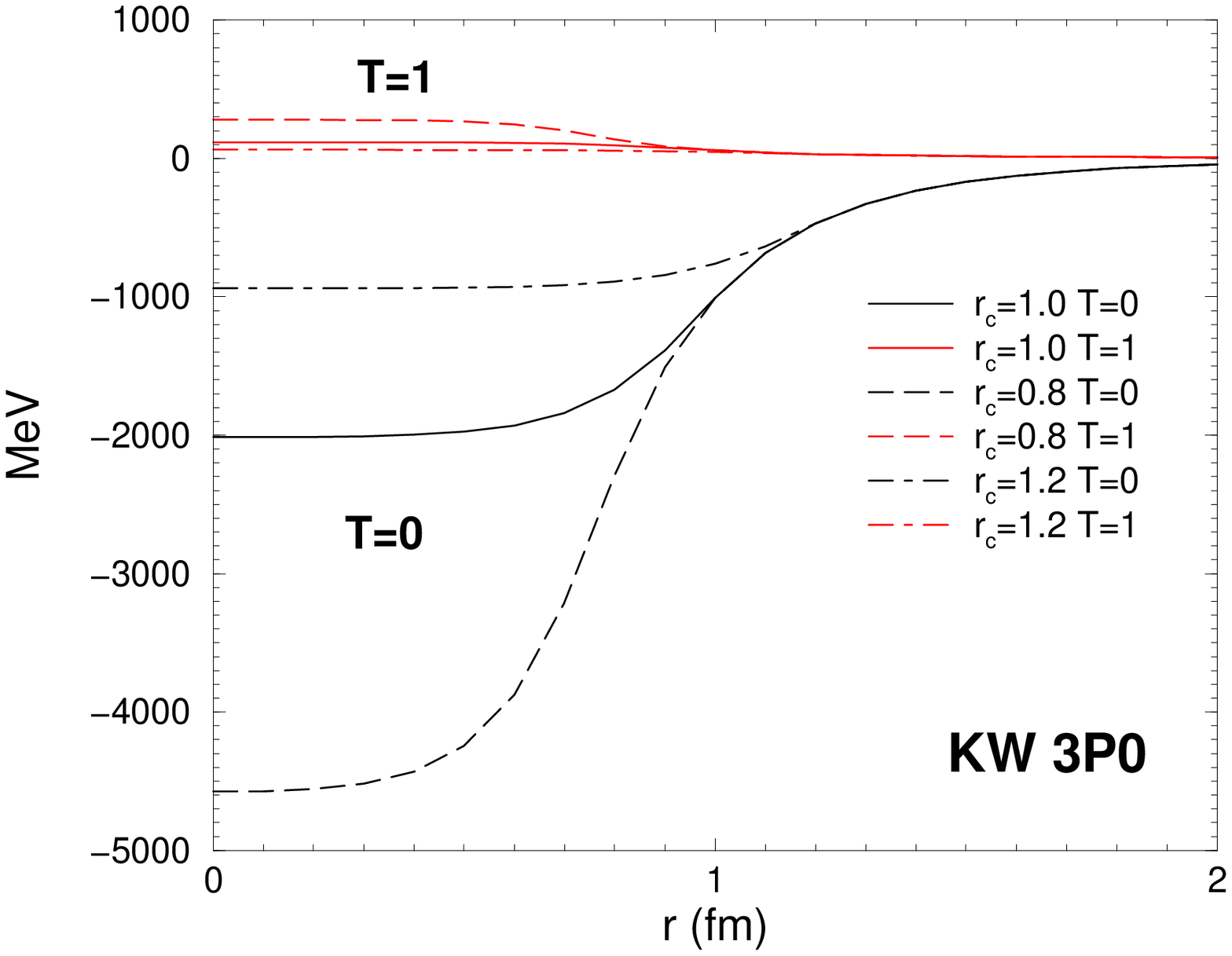}}
\vspace{-1.0cm}
\caption{V$_{\bar{N}N}$($^3P_0$)  in KW 
 model for different values of the cutoff parameter $r_c$.}\label{WKW_3P0_rc}
\end{minipage}
%\hspace{\fill}
\hspace{0.5cm}
\begin{minipage}[t]{75mm}
\begin{center}\epsfxsize=7.cm\mbox{\epsffile{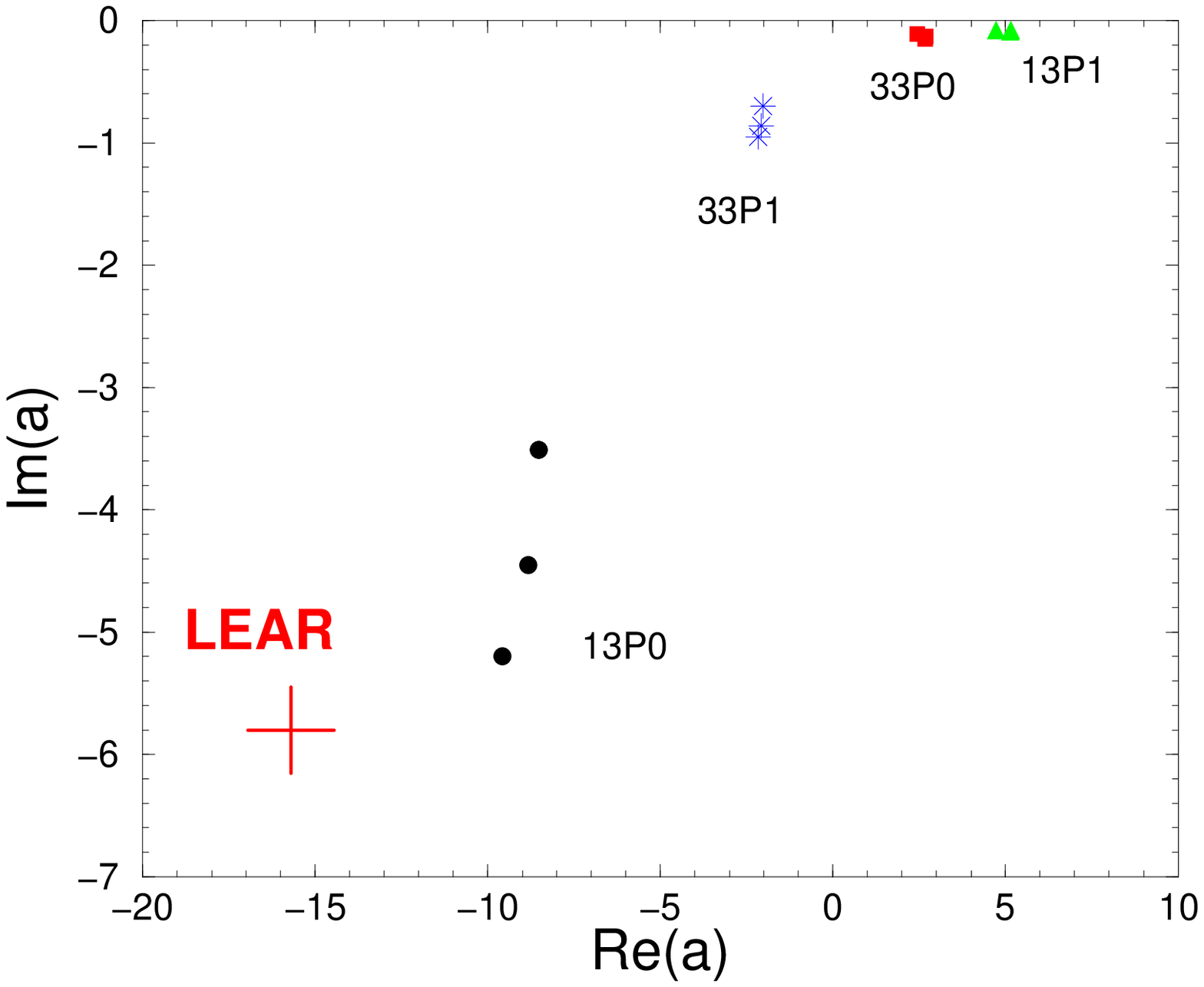}}\end{center}
\vspace{-1.0cm}
\caption{Isospin components for \={N}N scattering volumes in KW and DR models.}\label{LDIF_NNB_Gotta}
\end{minipage}
\vspace{-0.2cm}
\end{figure}

The predictions from KW \cite{KW_NPA_86}, DR1 and DR2 \cite{DR 80} optical models for 
the protonium P-level shifts \cite{CIR_ZPA_89} 
together with the corresponding \={p}p scattering volumes \cite{CRW_ZPA_92}
are given in Table \ref{POM}.
These models differ from each other both in the annihilation potential 
and in the meson contents defining its real part. 
One can see from this table that the $^3P_0$ width is well reproduced  
and that the energy shift
is -- in spite of the large predicted values -- underestimated by a factor two.   
We would like to emphasize however that
the very prediction of these models is that $^3P_0$ must be a particular case.
The reason for $\Delta E_R$ being 
2-10 times bigger than other P states
is that V$_{^3P_0}$ has an unusually strong and attractive T=0 component
at distances at which the OBEP theory should be reliable (see Figure \ref{WKW_3P0_rc}).
This large value cannot be reproduced by simple interactions: 
Bruckner et al. potential, for instance, 
which proved to be very successful at higher energy \cite{B_PLB166_86},  gives 
$a_{sc}$=$-0.18-0.68\,i$ fm$^3$.
This non trivial result seems be confirmed by experiment 
and -- despite a factor two in $\Delta E_R$ -- gives support to meson exchange inspired models.

%%%%%%%%%%%%%%%%%%%%%%%%%%%%%%%%%%%%%%%%%%%%%%%%%%%%%%%%%%%%%%%%%%%%%%%%%%%%%%%%%%%%%%%%%%%%%%%%%%%%%%

A little bit more hidden but even more interesting is the
way in which the P-wave \={p}p scattering volumes are obtained
in terms of the \={N}N  isospin components.
For that purpose we consider the DR1 prediction from Table \ref{POM}.
This value, denoted by $a_{sc}=-3.01-2.31\,i$, was calculated
using \={p}p-\={n}n coupled equations in which Coulomb ($V_c$) 
and proton-neutron mass difference ($\Delta$) were included.
By removing $\Delta$ and Coulomb corrections,
one is left with a purely strong value $a_s$, which in its turn
is an average of both isospin components $a_s={1\over2}\left(a_0+a_1\right)$. 
The result of this analysis, performed in \cite{CRW_ZPA_92}  for all partial amplitudes, 
is summarized in Table \ref{T_components}.

\begin{table}[htb]
\caption{Isospin components for the $^3P_0$ scattering volume}\label{T_components}
\newcommand{\m}{\hphantom{$-$}}
\newcommand{\cc}[1]{\multicolumn{1}{c}{#1}}
\renewcommand{\tabcolsep}{0.5pc} % enlarge column spacing
\renewcommand{\arraystretch}{1.} % enlarge line spacing
\begin{tabular}{llrrr}\hline
          &                  & DR1 \phantom{ab} &\phantom{ab}  & Exp.\phantom{cccccccccab}   \\\hline
$a_{sc}$  &                  &-3.01 -2.31 i  &\phantom{ab}  &  (-5.68$\pm$1.14)  -  (2.45$\pm$0.53) i \phantom{$^{(*)}$} \\
%         & $V_C$=0          &-3.32 -2.38 i  &              &                           \\
$a_s$     & $V_C$=$\Delta$=0 &-3.44 -2.66 i  &              &  (-6.5$\pm$1.3)  - (2.8$\pm$0.6) i   $^{(*)}$ \\
$a_{0}$   &                  &-9.58 -5.20 i  &              &   (-15.7$\pm$2.6) - (5.8$\pm$1.2) i  $^{(*)}$ \\
$a_{1}$   &                  &+2.69 -0.13 i  &              &             +2.69 -           0.13 i $^{(*)}$\\\hline
\end{tabular}
\\[2pt]
$^{(*)}$ obtained by theoretical analysis.
\vspace{-0.5cm}
\end{table}

By applying the same kind of corrections to the experimental 
scattering volume $a_{\bar{p}p}=-5.68-2.45\,i$ fm$^3$  and  
keeping the same T=1 value (this approximation
will be justified later) one gets  $a_{0}\approx-15.7-5.77\,i$.
The large value of its real part is a consequence 
of the already large measured Re$(a_{\bar{p}p})<0$
and the destructive interference with the repulsive T=1 channel.
It cannot be obtained without a
nearthreshold \={N}N state which enhances the scattering amplitude.
To our opinion, the results of \cite{G_NPA_99} 
constitutes a direct evidence for its existence.
It is interesting to compare this situation
with the singlet deuteron channel  $a_{np}\approx -20$ fm which inspired
the following conclusion:
"If deuteron had not been found experimentally one could infer
its very existence by the big values of NN scattering parameters".
Does G-parity manifest itself with such a degree of refinement ?

In Figure \ref{LDIF_NNB_Gotta} we have compared the different isospin components
of the \={N}N P states as they are given in \cite{CRW_ZPA_92} by the three considered optical models. 
One can see there, that if theoretical predictions for $^3P_0$ were not "normal",
the experimental findings at LEAR go well beyond. 
It is thus interesting to inquire
whether or not the  existing models are able to reproduce them.

\begin{figure}[htb]
\vspace{-0.3cm}
\begin{minipage}[t]{75mm}
\hspace{0.4cm}\epsfxsize=7.0cm\mbox{\epsffile{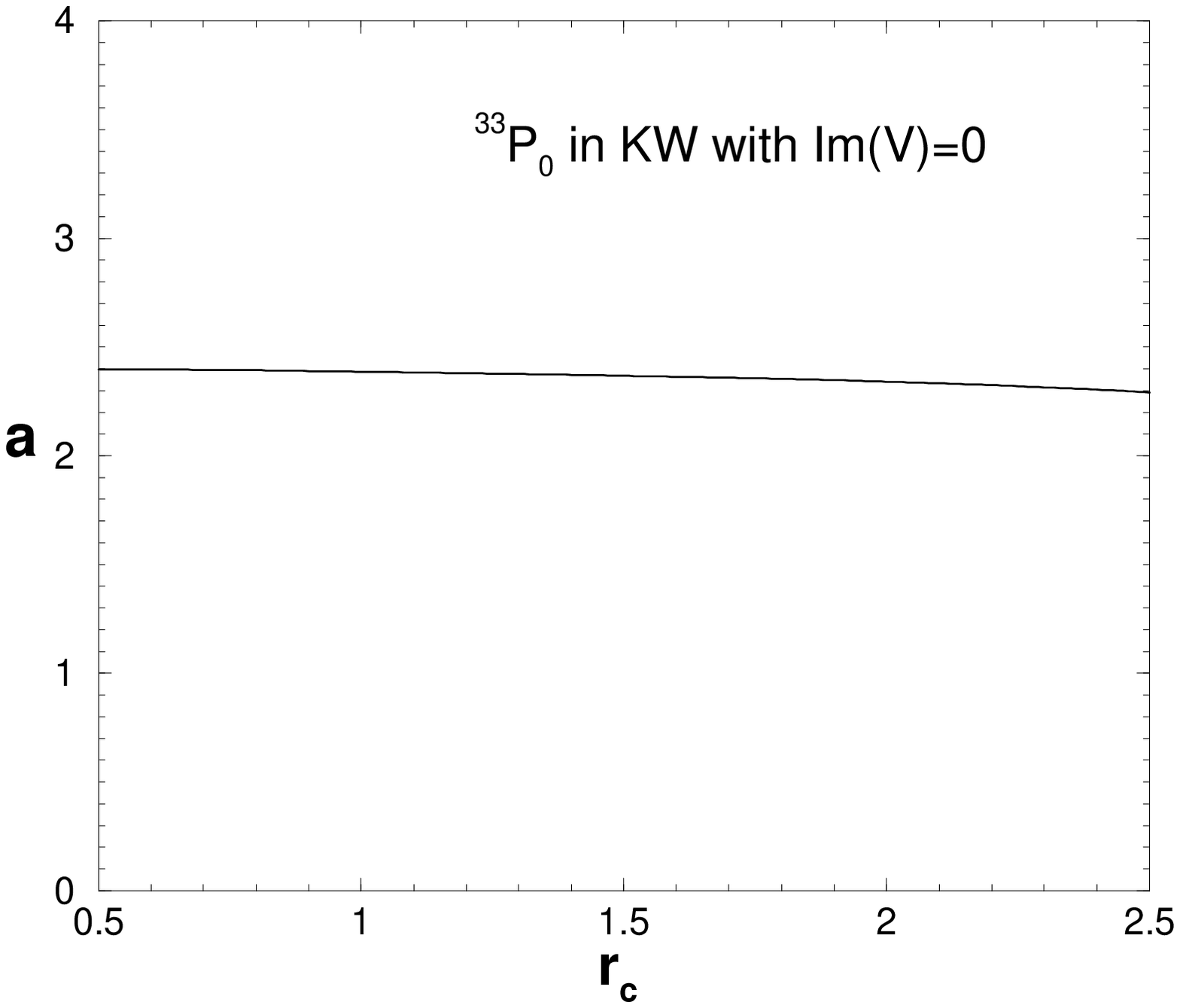}}
\vspace{-1.0cm}
\caption{$^{33}P_0$ scattering volume in KW  with Im(V)=0 as a function of $r_c$.}
\label{a_rc_33P0_ReKW}
\end{minipage}
\hspace{0.5cm}
\begin{minipage}[t]{75mm}
\hspace{0.4cm}\epsfxsize=7.0cm\mbox{\epsffile{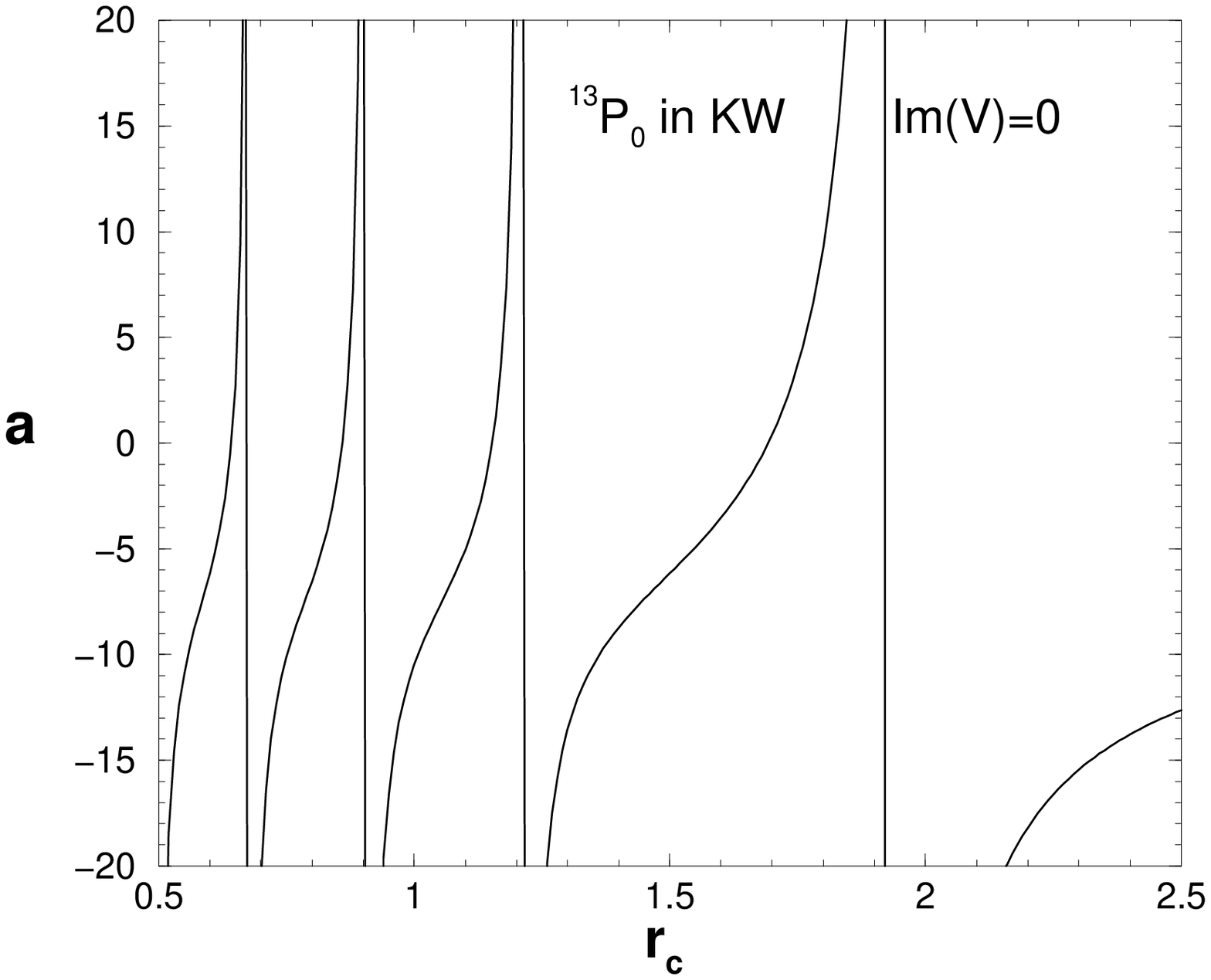}}
\vspace{-1.0cm}
\caption{$^{13}P_0$ scattering volume in KW  with Im(V)=0 as a function of $r_c$.}
\label{a_rc_13P0_ReKW}
\end{minipage}
\vspace{-0.5cm}
\end{figure}

Let us first consider the problem with Im(V)=0 and  
calculate the $^3P_0$ scattering volume $a$ as a function of the strength of Re(V), i.e. the cutoff radius $r_c$.
The results for both isospin components are  plotted in Figures \ref{a_rc_33P0_ReKW} and \ref{a_rc_13P0_ReKW}.
The  $^{33}P_0$ state shows a remarkable stability due to the fact that the corresponding
potential -- displayed in Figure \ref{WKW_3P0_rc} -- is 
repulsive and dominated by the centrifugal barrier. 
For the  $^{13}P_0$  state, $a$ varies from $-\infty$ to $+\infty$ -- what makes easy 
fitting any experimental result --
the vertical asymptotics corresponding to the appearence of new \={N}N bound states.
One can also see from these figures that the negative sign of the 
scattering volume ($a<0$) implies attractive interaction 
whereas from $a>0$ nothing can be infered.
%%%%%%%%%%%%%%%%%%%%%%%%%%%%%%%%%%%%%%%%%%%%%%%%%%%%%%%%%%%%%%%%%%%%%%%%%%%%%%%%%%%%%%%%%%%%%%%%%%%%%%
\begin{figure}[htb]
\vspace{-0.0cm}
\begin{minipage}[t]{75mm}
\hspace{0.3cm}\epsfxsize=7.cm\mbox{\epsffile{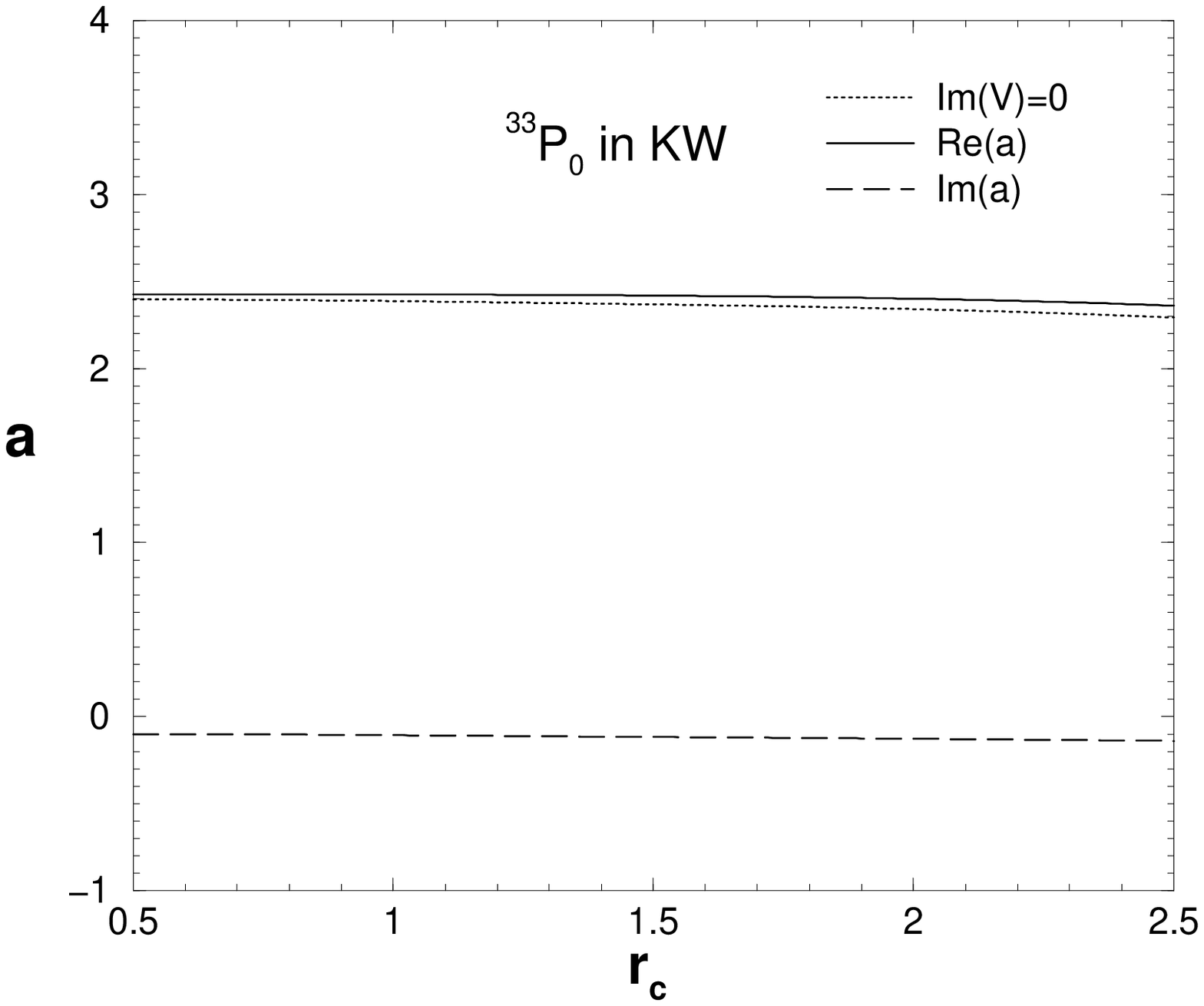}}
\vspace{-1.1cm}
\caption{$^{33}P_0$ scattering volume in KW as a function of the cutoff 
parameter $r_c$.}
\label{a_rc_33P0_KW}
\end{minipage}
\hspace{0.5cm}
\begin{minipage}[t]{75mm}
\hspace{0.3cm}\epsfxsize=7.cm\mbox{\epsffile{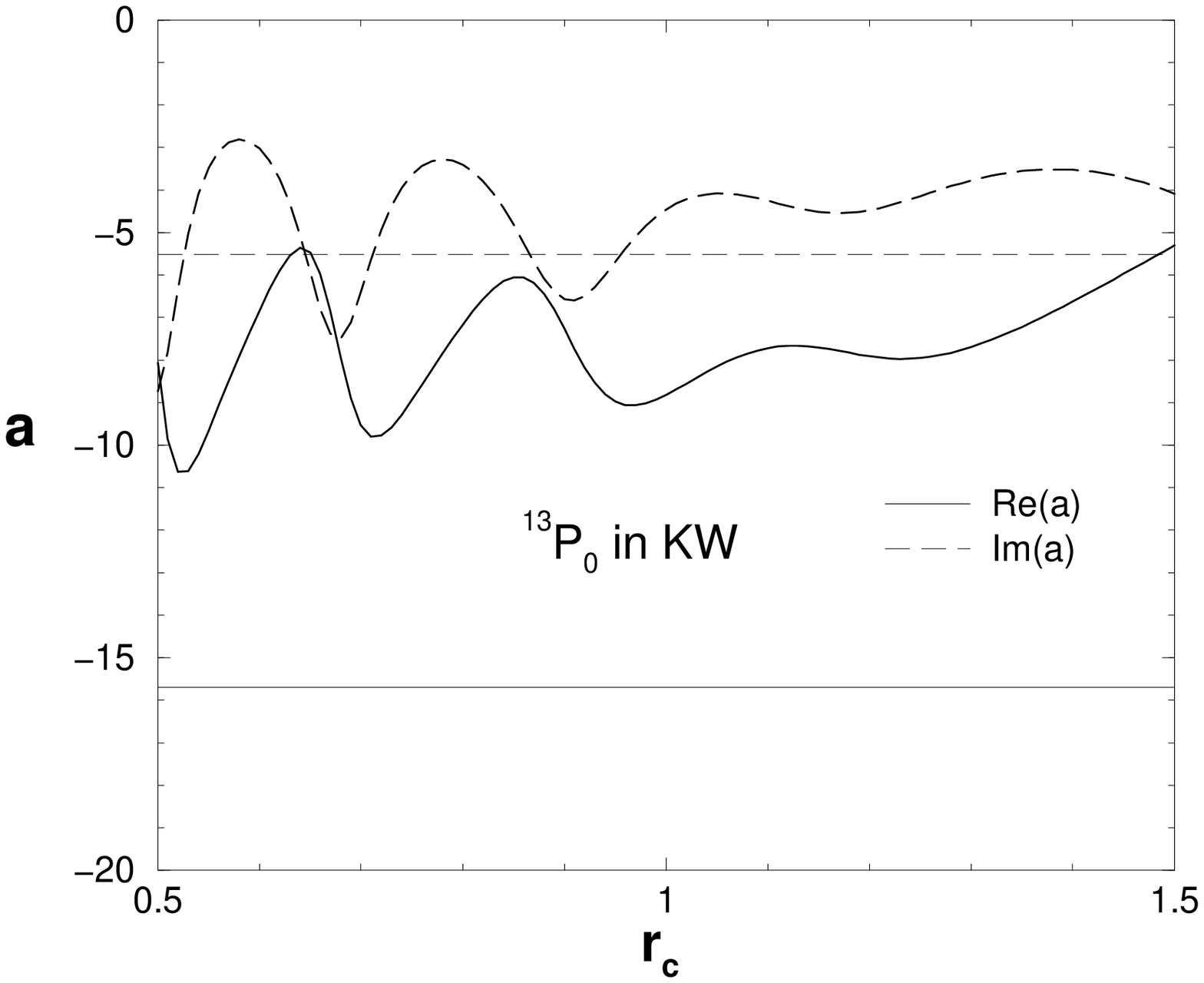}}
\vspace{-1.1cm}
\caption{$^{13}P_0$ scattering volume in KW as a function of the cutoff parameter $r_c$.}
\label{a_rc_13P0_KW}
\end{minipage}
\vspace{-0.5cm}
\end{figure}

\begin{figure}[htb]
\vspace{-0.3cm}
\begin{minipage}[t]{75mm}
\hspace{0.3cm}\epsfxsize=7.cm\mbox{\epsffile{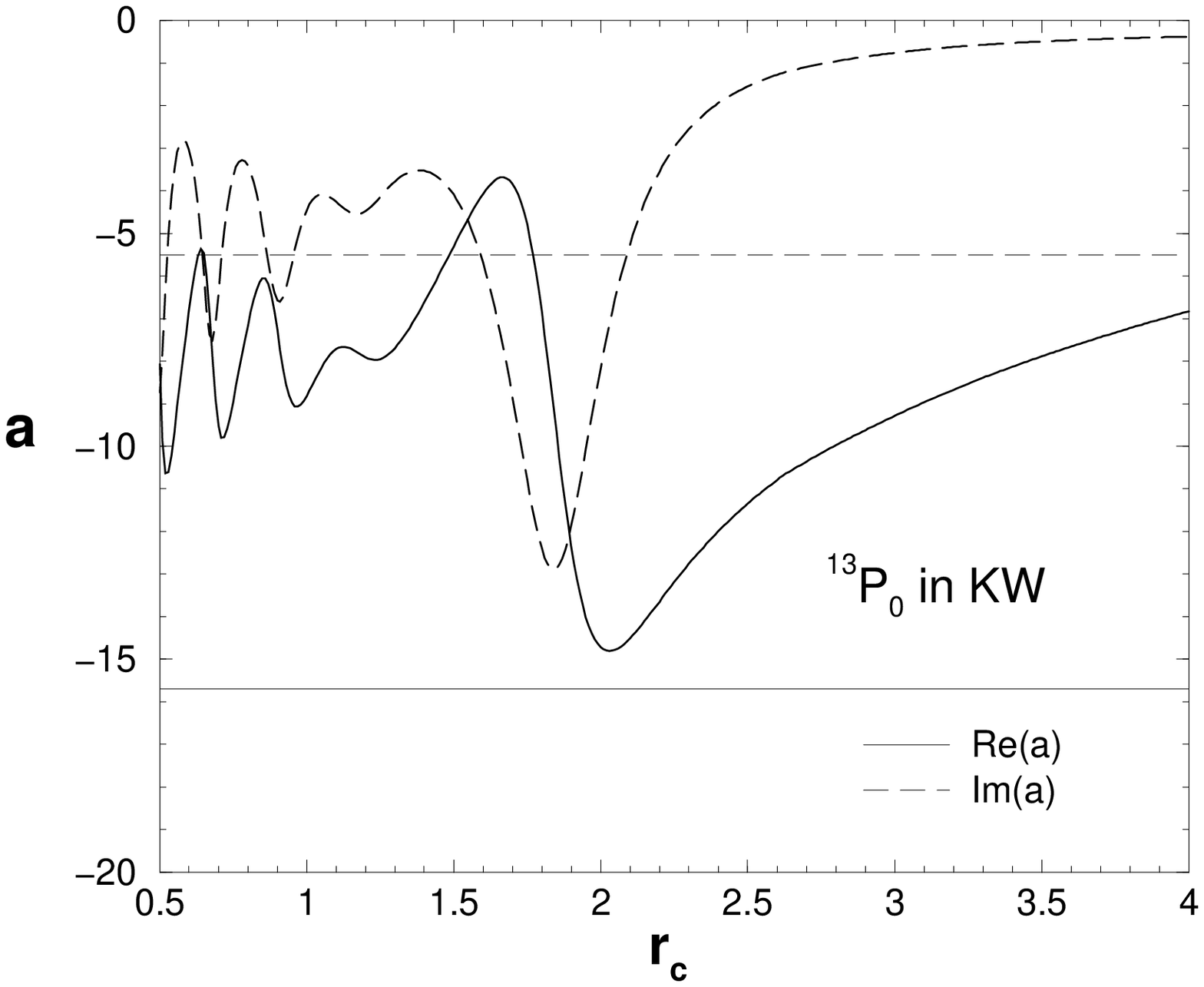}}
\vspace{-1.0cm}
\caption{Solution for large values of the cutoff parameter $r_c$.}\label{arc_13P0_KW_4.0}
\end{minipage}
\hspace{0.5cm}
\begin{minipage}[t]{75mm}
\hspace{0.3cm}\epsfxsize=7.cm\mbox{\epsffile{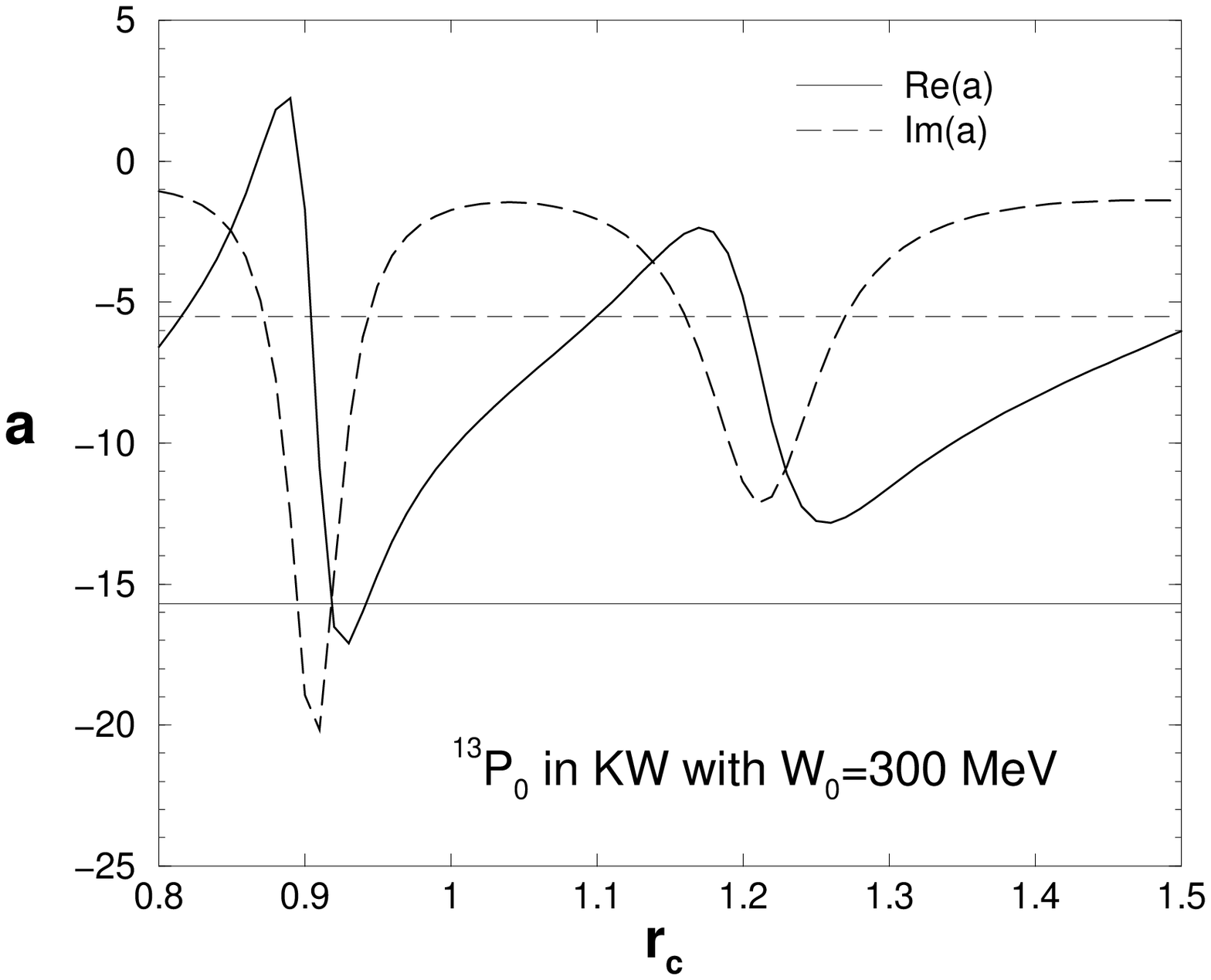}}
\vspace{-1.0cm}
\caption{Solution with Im(V)=300 MeV and 
usual values of the cutoff $r_c$.}\label{arc_13P0_0.25ImKW_2}
\vspace{-0.7cm}
\end{minipage}
\end{figure}

In presence of annihilation, the results for the $^{33}P_0$ state are only slightly modified. 
For a wide range of $r_c$ values, $a$ keeps nearly the same real part and 
gets only a very small imaginary part, indicating that 
annihilation cannot come from repulsive channels.
On the contrary the behaviour of the $^{13}P_0$ is much more complex.
The pole divergences are reduced to finite oscillations and $a$ can have large imaginary part.
In order to agree with the experimental values, Re(a) and Im(a)
should cross the corresponding horizontal line at the same value of $r_c$.
If one takes the results from \cite{G_NPA_99} this is not possible even modifying
the cutoff parameter in a reasonable range.
We have found several solutions to account for that 
interesting -- though disappointing -- situation.
The first one, see Figure \ref{arc_13P0_KW_4.0}, keeps the same annihilation potential 
but uses a very large value of $r_c$=2 fm. The corresponding
potential becomes only of $\sim$100 MeV but we notice 
that even in this case there is a quasibound state and the 
experimental data force the model to be in its vicinity.
The second solution is obtained  by a substantial change in Im(V),
thus introducing a quantum number dependence in annihilation potential, 
as it was the case in Paris potential \cite{CLLMV_PRL_82}. 
Figure \ref{arc_13P0_0.25ImKW_2} shows a solution obtained 
with a factor 4 reduction in the strength $W_0$ of the imaginary potential 
($W_0$=300 MeV instead of 1.2 GeV) and $r_c=0.93$ fm.
 
%%%%%%%%%%%%%%%%%%%%%%%%%%%%%%%%%%%%%%%%%%%%%%%%%%%%%%%%%%%%%%%%%%%%%%%%%%%%%%%%%%%%%%%%%%%%%%%%%%%%%%

To conclude  this section, we would like to emphasize the importance
of low energy experiments as well as
the partial wave resolution to get useful information about the interaction.
The existence of $^{33}P_0$ state cannot directly by deduced from the total
\={p}p scattering cross section due to its small statistic weight
compared to the large number of partial waves involved.
In addition, it appears only in the T=0 part
of the \={p}p wavefunction and its influence is partially cancelled by the 
opposite
contribution of the T=1 part. It is worth mentioning however that a similar conclusion was
also reached in \cite{CDPS_NPA_91} based on an independent analysis.

%%%%%%%%%%%%%%%%%%%%%%%%%%%%%%%%%%%%%%%%%%%%%%%%%%%%%%%%%%%%%%%%%%%%%%%%%%%%%%%%%%%%%%%%%%%%%%%%%%%%
\section{Selected topics on \={N}d}

The interest in studying this process -- apart from testing \={N}N models in a more
complex system --  is to have an idea of how  \={N}N annihilation takes place in presence of other nucleons.
Understanding annihilation at a quark level is made difficult by the fact that
we have no appropriate theoretical tools to deal with quarks (q).
Calculations are performed on a basis of constituent quarks (Q), mostly in a non relativistic approach.
However Q is a very complex object which looks more like N than  q
except that it has no asymptotic states.

In \={N}d, we know something -- at least experimentally -- about the elementary amplitudes
and it seems in principle easier to inquire 
about the \={N}-A annihilation mechanism.
Does  \={N} annihilate on a single N ?
If yes, should we expect $T_{\bar{p}A} \approx T_{\bar{p}N}$ ? 
or rather $T_{\bar{p}A}\approx T_{\bar{p}N_1}+T_{\bar{p}N_2}$?
and what is the role of the remaining nucleons ?  
Does on the contrary \={N} annihilate on the nuclear bulk ?
From that purpose deuteron -- quite an extended object --
is the most interesting nucleus 
and low energy  experiments are the only ones from which one
 can get useful information.
The problem has some analogies with 
"From where does the e$^-$ pass through ? " in a Young experiment,
where we learnt that naive  pictures (and questions!) have hard 
life in Quantum Mechanics.

\={N}d is a genuine problem for the Few-Body Physics community. 
Faddeev-Yakubovsky (FY) equations in configuration space seem
 to be the best adapted to deal
with non hermitic problems where Coulomb interactions play a vital role.
At present, the more complex system for which FY 
can be exactly solved is N=4 \cite{CC_PRC_98}. 
The only \={N}A problem that can be presently solved is \={N}-d. It
could be extended in coming years to \={N}-$^3$He and 
-- if no bad surprises -- to \={N}-$^4$He. A modest but already fascinating task.

From nuclear physics, we know that the scattering lengths are very dynamical quantities, i.e.
very sensible to and determined by the interaction.
Any interpretation in terms of geometry is purely lyric. 
For instance, the singlet n-p value is $a^{(1)}_{np}$=--23 fm for 
a nucleon  r.m.s. radius of R=0.86 fm.
On the other hand, the doublet and quartet n-d are respectively
$a^{(2)}_{nd}$=0.65 fm and $a^{(4)}_{nd}$=6.4 fm,
i.e. they concern the same geometrical object and differ by one order of magnitude. 
%Geometry appears at high energy (lot of cancelling contributions ).
One should anyway remind that the size of a light
nucleus is mainly determined by its binding energy rather than by the
number of constituents: deuteron is  bigger than $\alpha$ particle.
This sensitivity to details of the interaction made powerful approximate methods 
 fail at very low energies.
Only well founded approaches that take into account the 
full dynamics produce there reliable results.

If the low energy parameters are dynamical, 
one cannot exclude some \={N}A states having  very weak annihilation rates.
An example -- concerning pd elastic scattering -- 
is given by  $a_{pd}^{(2)}\approx$0 fm ! 
In \={p}p, one has 
--  according to theoretical predictions \cite{CRW_ZPA_92}) -- 
 Im($a_{T=0}$)=0.08 fm$^3$ for $^3$P$_1$. 

The results that follow have been obtained by solving the Faddeev equations
with the methods developed for
N=3 \cite{CGM_FBS_93}  and N=4 \cite{CC_PRC_98}. 
In the Faddeev approach,
 the total \={n}d wave function is obtained as a sum of three amplitudes:
\begin{equation}\label{Psi}
\Psi= \Psi_{pn}(x_1,y_1) + \Psi_{\bar{n}p}(x_2,y_2) + \Psi_{\bar{n}n}(x_3,y_3)
\end{equation}
each of them depending on a particular set of Jacobi coordinates:
\begin{figure}[hbtp]
\vspace{-0.3cm}
\begin{center}\epsfxsize=11.0cm\mbox{\epsffile{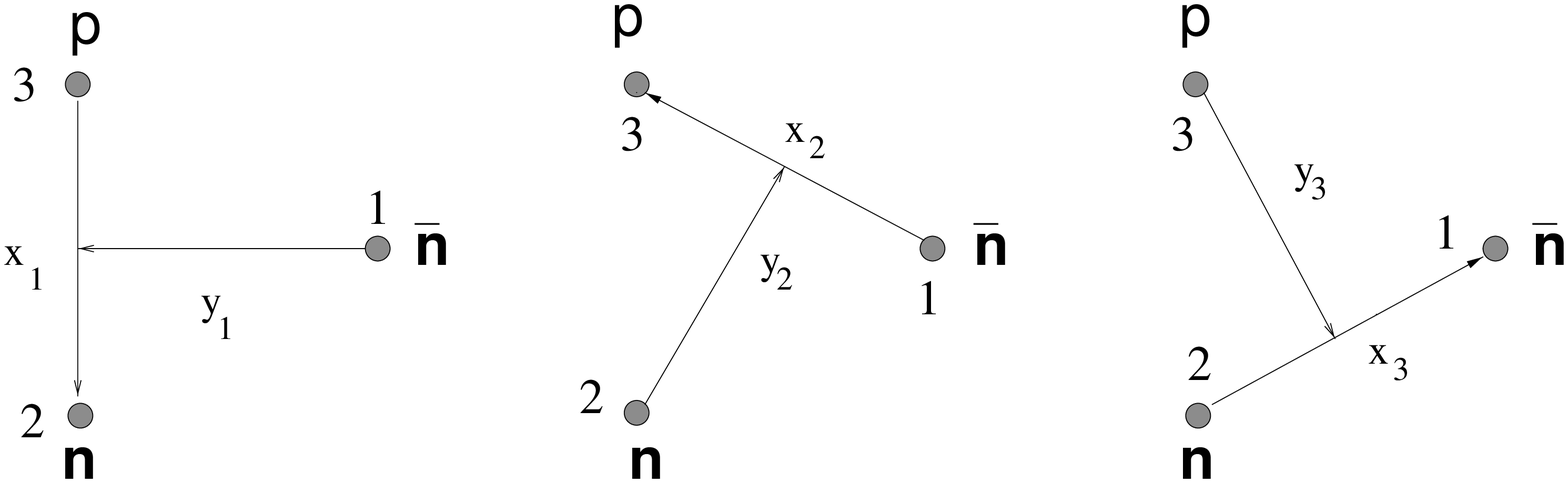}}\end{center}
\vspace{-0.8cm}
\end{figure}

These amplitudes obey a system of coupled equations 
\begin{eqnarray*}
(E-H_0-V_{pn})       \Psi_{pn}       &=& V_{pn}       ( \Psi_{\bar{n}p} + \Psi_{\bar{n}n} )  \cr
(E-H_0-V_{\bar{n}p}) \Psi_{\bar{n}p} &=& V_{\bar{n}p} ( \Psi_{\bar{n}n} + \Psi_{pn}       )  \cr
(E-H_0-V_{\bar{n}n}) \Psi_{\bar{n}n} &=& V_{\bar{n}n} ( \Psi_{pn}       + \Psi_{\bar{n}p} ) 
\end{eqnarray*}
and each of them is in its turn expanded in partial wave components
\[  \Psi_{i} = \sum_{\alpha_i} \Psi_{i,\alpha_i} \qquad \alpha_i= \{l_{x_i},\sigma_{x_i},j_{x_i},l_{y_i},j_{y_i}\} \]
on which the different partial waves of the NN and \={N}N potentials act.
Only each separate component has a well defined asymptotic behaviour which, for 
the scattering problem we are interested in, is:
\begin{eqnarray*}
\Psi_{pn}      (x_1,y_1) \approx u_{pn}(x_1)       \left[ e^{-iq_1y_1} - {\boldmath  S_{11}}\, e^{-iq_1y_1} \right] \cr
\Psi_{\bar{n}p}(x_2,y_2) \approx u_{\bar{n}p}(x_2) \left[ \phantom{e^{-iq_2y_2}} - {\boldmath  S_{12}}\, e^{-iq_2y_2}\right ] \cr
\Psi_{\bar{n}n}(x_3,y_3) \approx u_{\bar{n}n}(x_3) \left[ \phantom{e^{-iq_3y_3}} - {\boldmath  S_{13}}\, e^{-iq_3y_3}\right]
\end{eqnarray*}
where $u_{i}(x_i)$ is the \={N}N two-body bound state wavefunction 
with energy $E_i={2\over3}{q_i^2\over M}$ and $S_{ij}$ the S-matrix elements.
For a given \={N}N model, the values of $E_i$ depend on 
the quantum numbers \{$l_x\sigma_xj_x$\} of the \={N}N state.\par
In practice, all models have 
one or several \={N}N quasibound states with energies below the deuteron.
The spectrum of KW model, for instance, is displayed in Table \ref{Spectre_KW} 
for both isospin components of the potential
as well as for their linear combination $V_{\bar{n}n}={1\over2}(V_0+V_1)$
which governs the \={n}n channel. One can remark that there are no quasibound 
states for \={n}p (T=1) but two  in \={n}n. 
\begin{table}[htb]
\newcommand{\m}{\hphantom{$-$}}
\newcommand{\cc}[1]{\multicolumn{1}{c}{#1}}
\renewcommand{\tabcolsep}{1.2pc} % enlarge column spacing
\renewcommand{\arraystretch}{1.} % enlarge line spacing
\caption{Quasibound states in KW model (MeV)}\label{Spectre_KW}
\begin{tabular}{l rrr}\hline  
State     & $V_0$       & $V_1$    & $V_{\bar{n}n}$      \\  \hline
$^1S_0$   & ---         & ---   & ---       \\ 
$^3SD_1$  & -965-438i    & ---   &	-109-378i\\
          & -76-374 i   & ---   & ---	 \\ 
$^3P_0$   & -1155-465i   & ---   & -243-346i  \\
          & -213-311i   &         & ---    \\
$^3PF_2$  & -619-338i    & ---    & ---       \\\hline
\end{tabular}
\vspace{-0.3cm}
\end{table}
Even if the initial state has only one asymptotics, e.g. deuteron, 
the coupled equations will populate all the others and
there will be always a non zero probability to leave the initial channel.
As a consequence, even in absence of annihilation potential, 
the \={n}d scattering length will have an imaginary part.
We can thus distinguish two kinds of annihilation:
the direct one and
the one that takes place after the creation of a quasibound \={N}N state.
In a practical calculation we cannot avoid 
taking into account these rearrangement channels. 

The results we have obtained 
for the two \={n}d S-wave states -- doublet ($J^{\pi}={1\over2}^+$) and 
quartet ($J^{\pi}={3\over2}^+$)  -- are respectively 
$a^{(2)}_{\bar{n}d}$=$1.59-0.88\,i$ fm and 
$a^{(4)}_{\bar{n}d}$=$1.62-0.82\,i$ fm, what gives an spin-averaged value
$\bar{a}_{\bar{n}d}$=$1.61-0.84\,i$ fm.
Each \={N}N Faddeev amplitude involves the $\{l_{x_i},\sigma_{x_i},j_{x_i},l_{y_i},j_{y_i}\}$
components  listed in Table
\ref{FADC} which include all $V^{j\leq1}_{\bar{N}N}$ interactions.

It is interesting to compare these results to those given by the
\={N}N amplitudes which are relevant in \={n}d. 
Their spin-averaged values, taken from Table 1 in \cite{CRW_ZPA_92}, are  
$\bar{a}_{\bar{n}n}$=$0.88-0.84\,i$ fm and $\bar{a}_{\bar{n}p}$=$0.85-0.76\,i$ fm.
One then has Im($\bar{a}_{\bar{n}d}$)$\ge$Im($\bar{a}_{\bar{N}N}$),
in contrast to  the findings of \cite{PBLZ_EPJA_00} for \={p}d.
In view of this results one is tempted to conclude that \={n}d
annihilation is dominated by the \={n}n channel.
However in order to get some light on the annihilation mechanism an analysis of the \={n}d wavefunction
is required.

\begin{table}[htb]
\vspace{-0.5cm}
\newcommand{\m}{\hphantom{$-$}}
\newcommand{\cc}[1]{\multicolumn{1}{c}{#1}}
\renewcommand{\tabcolsep}{0.9pc} % enlarge column spacing
\renewcommand{\arraystretch}{1.} % enlarge line spacing
\caption{Faddeev components involved in calculating \={N}d scattering lengths}\label{FADC}
\[\begin{array}{|ll|ll|c|c|} 
\multicolumn{6}{c}{J^{\pi}=1/2^+}\\ \hline
j_x& j_y &\sigma_x&l_x& V     & l_y      \\ \hline
0  & 1/2 & 0      & 0 & ^1S_0 & 0  	     \\ 
   &     & 1      & 1 & ^3P_0 & 1  	     \\\hline
1  & 1/2 & 0      & 1 & ^1P_1 & 1        \\
   &     & 1      & 0 & ^3S_1 & 0        \\
   &     & 1      & 1 & ^3P_1 & 1        \\
   &     & 1      & 2 & ^3D_1 & 0        \\ \hline
   & 3/2 & 0      & 1 & ^1P_1 & 1        \\ 
   &     & 1      & 0 & ^3S_1 & 2        \\  
   &     & 1      & 1 & ^3P_1 & 1        \\  
   &     & 1      & 2 & ^3D_1 & 2        \\  \hline
\multicolumn{6}{c}{ }\\       
\multicolumn{6}{c}{ }\\       
\multicolumn{6}{c}{ }\\       
\multicolumn{6}{c}{ }\\       
\end{array} 
\qquad
\begin{array}{| ll | llr |c |} 
\multicolumn{6}{c}{J^{\pi}=3/2^+}\\ \hline
j_x& j_y &\sigma_x&l_x& V        & l_y      \\ \hline
0  & 3/2 & 0      & 0 & ^1S_0   & 2          \\ 
   &     & 1      & 1 & ^3P_0   & 1           \\\hline
1  & 1/2 & 0      & 1 & ^1P_1   & 1         \\
   &     & 1      & 0 & ^3S_1   & 0         \\
   &     & 1      & 1 & ^3P_1   & 1         \\
   &     & 1      & 2 & ^3D_1   & 0        \\ \hline
   & 3/2 & 0      & 1 & ^1P_1   & 1         \\ 
   &     & 1      & 0 & ^3S_1   & 2        \\  
   &     & 1      & 1 & ^3P_1   & 1         \\  
   &     & 1      & 2 & ^3D_1   & 2        \\  \hline
   & 5/2 & 0      & 1 & ^1P_1   & 3         \\ 
   &     & 1      & 0 & ^3S_1   & 2        \\  
   &     & 1      & 1 & ^3P_1   & 3        \\  
   &     & 1      & 2 & ^3D_1   & 2       \\  \hline
\end{array} \]
\vspace{-0.5cm}
\end{table}

\begin{figure}[htb]
\begin{minipage}[t]{75mm}
\hspace{0.3cm}\epsfxsize=7.cm\mbox{\epsffile{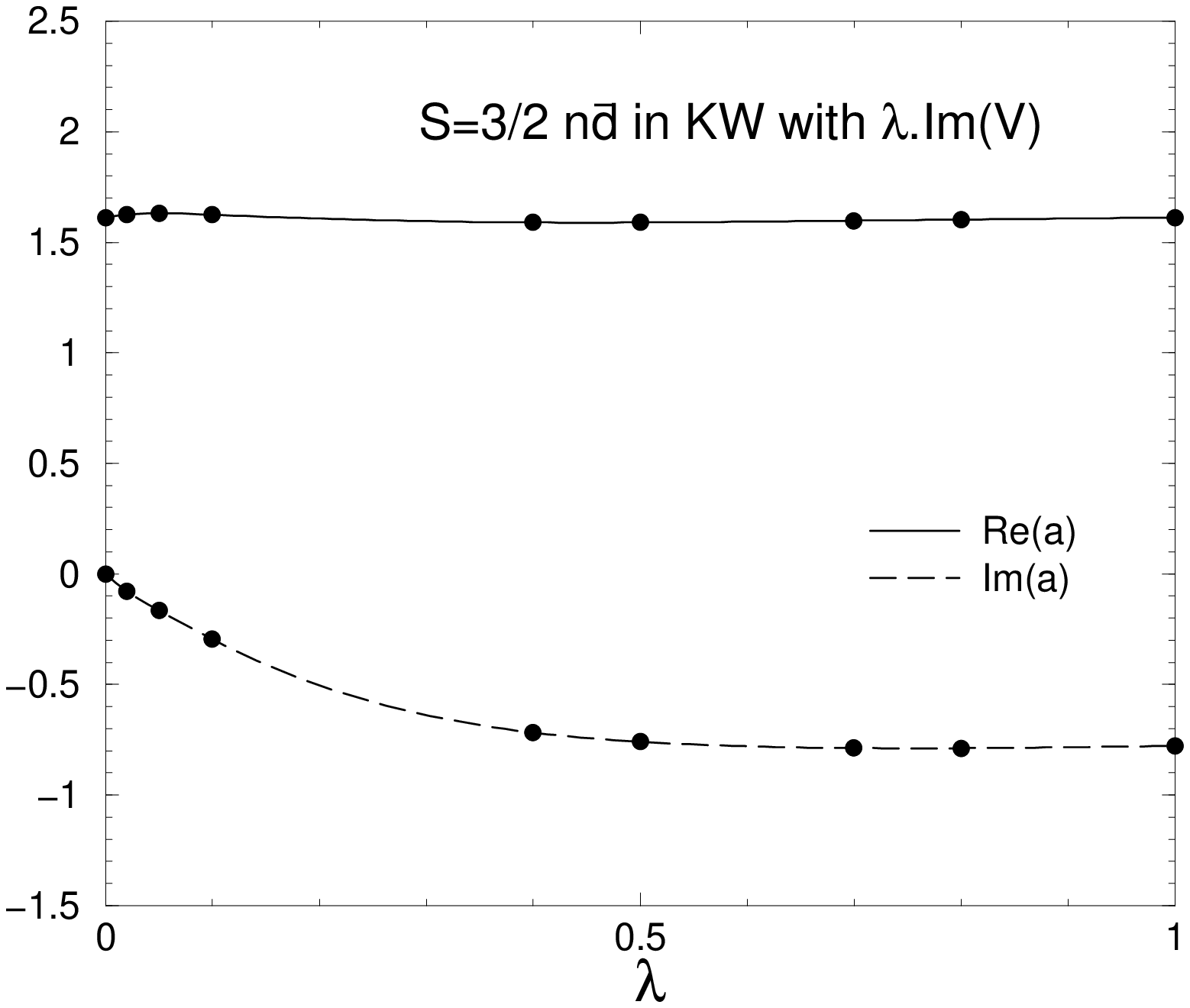}}
\vspace{-1.cm}
\caption{Evolution of $a_Q$ as a function of the strength of Im(V).}\label{aQ_W0}
\end{minipage}
\hspace{0.5cm}
\begin{minipage}[t]{75mm}
\hspace{0.3cm}\epsfxsize=7.0cm\mbox{\epsffile{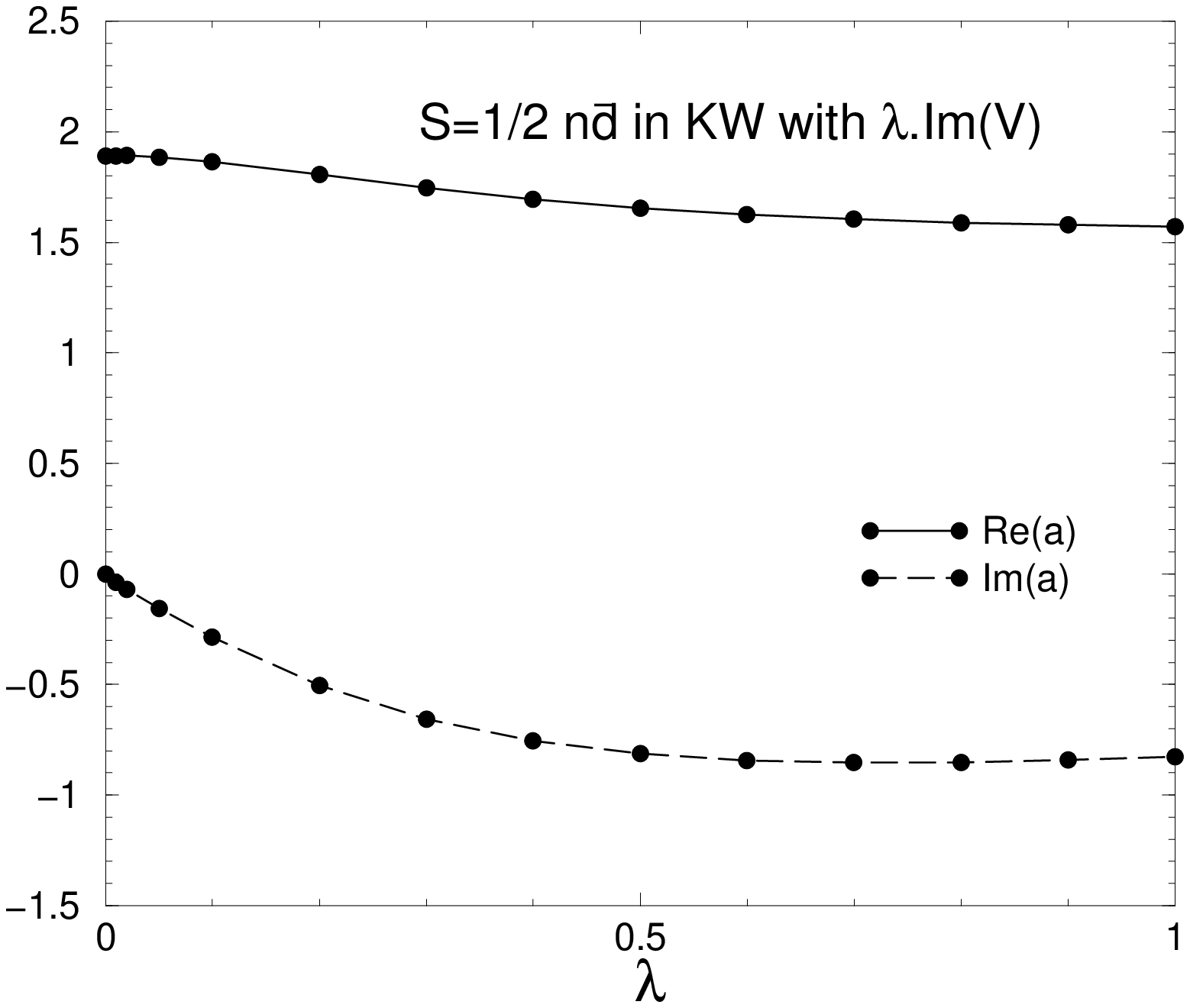}}
\vspace{-1.cm}
\caption{Evolution of $a_D$ as a function of the strength of Im(V).}\label{aD_W0}
\end{minipage}
\vspace{-0.3cm}
\end{figure}

We also remark that, contrary to nd case, there is very small spin dependence, that is $a^{(2)}\approx a^{(4)}$.
This could be expected from the fact that
we are dealing with three non identical particles.
It is worth noticing however
that with Im(V)=0, the values are $a^{(2)}\approx2.0$ and $a^{(4)}\approx1.5$.
The effect of annihilation potential is to move the two spin values 
in opposite directions towards an almost degenerate result. 
In Figures \ref{aQ_W0} and \ref{aD_W0} is displayed the variation
of the real and imaginary parts of $a_{\bar{n}d}$ as a function of a scaling
factor $\lambda$ multiplying Im(V).
They show a smoother behaviour than the one found in
the model calculation of \cite{KPV_EPJA_00}.

We note also that results can vary up to a factor two depending
on the states included in the rearrangement channels.
There are however dominant amplitudes which are $^1$S$_0$ and $^3$SD$_1$.
P-wave seems to account for only $\approx 10\%$ variation.

To our knowledge, there is no
any experimental value for \={n}d to compare with.
Wycech et al. \cite{WGN_PLB_85} found
$a_{2}=1.49-0.41 i$ and $a_{4}= 1.53-0.45 i$ and even smaller
values were found in \cite{LT_PRC_90}.
In view of comparing 
these results with those obtained in \cite{PBLZ_EPJA_00} from \={p}d experiment 
 -- Im($\bar{a}$)=0.62$\pm$0.02 -- Coulomb corrections should be included.

These calculations can be improved
in several ways. For instance by including D-wave component in deuteron, by
coupling the \={n}n amplitude to the \={p}p one and
by disentangling the contribution of the different rearrangement channels.
The sensitivity to asymptotic states
would also require to use different \={N}N models.
This work is being pursued by extending to \={N}d P-waves and
by including Coulomb interactions in the \={p}d system.

%%%%%%%%%%%%%%%%%%%%%%%%%%%%%%%%%%%%%%%%%%%%%%%%%%%%%%%%%%%%%%%%%%%%%%%%%%%%%%%%%%%%%%%%%%%%%%%%%%%%
\section{Conclusions}

Some years after its shutdown, LEAR continues to produce interesting results.
This shows that its unique characteristics -- low energy and high resolution
experiments --  were very appropriate for understanding the \={N}N interaction.
% The it is a very living community
If stopping LEAR was a mistake, it would be a fault to stop such an activity,
not adapting the AD facility to low energy scattering experiments,
against the request of several groups 
\cite{Felicello_LEAP_00,AM_LEAP_00,Gotta_PC}.

We have shown -- if needed -- the interest in pursuing low
energy antiproton physics in three different problems.
The intriguing low energy \={n}p observed structure could be also seen -- and thus confirmed --
in the elastic \={p}p cross section near the charge-exchange threshold.
The measured $^3P_0$ protonium level
shift constitutes a strong indication of a \={N}N nearthreshold state 
in isospin zero channel.
% In addition they force us to review the current optical models.
This measure should be confirmed and if possible
extended to other partial waves.
The \={N}d reaction is of interest in understanding the annihilation 
mechanism. We
have presented the first \={n}d results based 
on exact solutions of Faddeev equations.
This theoretical effort should be pursued by extending the calculations
to negative parity states 
-- where the $^{33}P_0$ pole could also manifest -- and to \={p}d system.

We would like to emphasize that only the knowledge of the isolated 
partial wave contributions provides effective constraints to models.
The $^3P_0$ constitutes a nice illustration of how
interesting phenomena, not even visible in the cross sections, 
can manifest after such a separation.
For that purpose -- and in absence of polarized beams -- 
the atomic experiments seem to be the only ones offering such a possibility.

\bigskip Acknowledgements.
The numerical calculations were performed  at CGCV (CEA Grenoble) and  IDRIS (CNRS).
We are grateful to the staff members of these two organizations for their constant support.

%%%%%%%%%%%%%%%%%%%%%%%%%%%%%%%%%%%%%%%%%%%%%%%%%%%%%%%%%%%%%%%%%%%%%%%%%%%%%%%%%%%%%%%%%%%%%%%%%%%%%%%


\begin{thebibliography}{99}
\bibitem{Iazzi_PLB_00}       F. Iazzi et al., Phys. Lett.  B475 (2000) 378      
\bibitem{Felicello_LEAP_00}  A. Felicello,   contribution to this Conference
\bibitem{Felicello_LEAP_98}  A. Felicello, Nucl. Phys. A655 (1998) 224 	 
\bibitem{Iazzi_LEAP_92}      A. Adamo et al., Nucl. Phys A558 (1993) 137c      %     The OBELIX Collaboration, Proceedings Courmayeur
\bibitem{Giacobbe_LEAP_96}   A. Bertin et al., Nucl. Phys. B 56A (1997) 227  	 %     The OBELIX Collaboration, Proceedings Dinkelskul
\bibitem{G_NPA_99}           D. Gotta et al, Nucl. Phys. A660 (1999) 283
\bibitem{KW_NPA_86}          M. Kohno and W. Weise, Nucl. Phys. A454 (1986) 429
\bibitem{DR 80}              C.B. Dover and J. M. Richard, Phys. Rev. C21 (1980) 1466
\bibitem{CIR_ZPA_89}	        J. Carbonell, G. Ihle, J.M. Richard,    Z. Phys. A 334 (1989) 329 
\bibitem{CRW_ZPA_92}         J. Carbonell, J. M. Richard, S. Wycech, Z. Phys. A 343 (1992) 325 
\bibitem{B_PLB166_86}        W. Bruckner et al., Phys. Lett. 166B (1986) 113  
\bibitem{CLLMV_PRL_82}       J. C\^ot\'e et al, Phys. Rev. Lett. (1982) 1319
\bibitem{CDPS_NPA_91}	       J. Carbonell, O. Dalkarov, K Protasov,I. Shapiro, Nucl.Phys. A535 (1991) 651
\bibitem{CC_PRC_98}          F. Ciesielski, J. Carbonell, Phys. Rev. C58 (1998) 58
\bibitem{CGM_FBS_93}         J. Carbonell, C. Gignoux, S.P. Merkuriev, Few-Body Syst. 15 (1993) 15
\bibitem{KPV_EPJA_00}        V. Karmanov, K. Protasov, A. Voronin, Eur. Phys. J. A8 (2000) 429
\bibitem{WGN_PLB_85}         S. Wycech, A.M. Green, J.A. Niskanen, Phys. Lett. B152 (1985) 308
\bibitem{LT_PRC_90}          G.P. Latta, P.C. Tandy, Phys. Rev. C42 (1990) 1207
\bibitem{PBLZ_EPJA_00}       K. Protasov, G. Bonomi, E. Lodi Rizzini, A. Zenoni, Eur. Phys. J. A7 (2000) 429
\bibitem{AM_LEAP_00}         A. Martin,   contribution to this Conference
\bibitem{Gotta_PC}           D. Gotta,   private communication
\end{thebibliography}
\end{document}